\documentclass[pra,superscriptaddress,showpacs,twocolumn]{revtex4-1}

\usepackage{graphicx}
\usepackage{amsmath}
\usepackage{amssymb}
\usepackage{bm}

\renewcommand{\Re}{\mathrm{Re}}
\renewcommand{\Im}{\mathrm{Im}}

\newcommand{\dd}{\mathrm{d}}
\newcommand{\ee}{\mathrm{e}}
\newcommand{\ii}{\mathrm{i}}
\newcommand{\tr}{\mathrm{tr}}
\newcommand{\calO}{\mathcal{O}}
\newcommand{\PT}{$\mathcal{PT}\,$}
\begin{document}
                    
%\title{The imaginary cubic potential:\\
%Summation of the perturbative series by order-dependent-mappings}

\title{Order-dependent mappings: strong coupling behaviour from\\ 
weak coupling expansions in non--Hermitian theories}

\author{Jean Zinn-Justin}

\email{Electronic mail: jean.zinn-justin@.cea.fr}

\affiliation{CEA, IRFU, Centre de Saclay, 91191 Gif-sur-Yvette Cedex, France\\
E-mail: jean.zinn-justin@.cea.fr}

\author{Ulrich D.~Jentschura}

\affiliation{Missouri University of Science and Technology, Rolla, 
Missouri 65409-0640, USA}

\begin{abstract}
A long time ago, it has been conjectured that a Hamiltonian with a potential of
the form $x^2+i v x^3$, $ v$ real, has a real spectrum. This conjecture has
been generalized to a class of so-called \PT symmetric Hamiltonians and some
proofs have been given. Here, we show by numerical investigation that the
divergent perturbation series can be summed efficiently by an order-dependent mapping
(ODM) in the whole complex plane of the coupling parameter $v^2$,
and that some information about the location
of level crossing singularities can be obtained in this way.
Furthermore, we discuss to which accuracy the strong-coupling limit can be
obtained from the initially weak-coupling perturbative expansion,
by the ODM summation method. The basic idea of the ODM summation method is the notion of order-dependent ``local'' disk of convergence and analytic continuation by an order-dependent mapping of the domain of analyticity augmented by the local disk of convergence onto a circle.
In the limit of vanishing local radius of convergence, which is the limit
of high transformation order, convergence is demonstrated both by
numerical evidence as well as by analytic estimates.
\end{abstract}


\maketitle

%
% Introduction
%
\section{Introduction}
\label{intro}
   
The order-dependent mapping (ODM) summation method has been introduced in
Ref.~\cite{rsezzin} and initially applied to series expansions of 
an integral whose perturbative expansion counts the number of  Feynman diagrams
with four-point vertices, to the quartic anharmonic oscillator, and to
renormalization group (RG) functions in the three-dimensional $\phi^4$ quantum
field theory \cite{rsezzin,rZJODM}. Later, other examples of quantum mechanics
and field theory type have been studied \cite{rLGZJ,rRGZJ,rJLKMPRR,rKPR,rKPRS}.
Some convergence proofs have been given in Refs.~\cite{rDuJo,rRGKKHS}.
However, these examples have in common, as a starting point, perturbations of
Hermitian quantum Hamiltonians.  In most of them  the ODM summation method
appears as a higher order extension of some variational calculation
\cite{rsezzin,rWECas}.  Therefore, we may wonder whether Hermiticity plays a role
in explaining some specific convergence properties of the method.

Here, we thus consider a non-Hermitian example, the  expansion generated by an
$\ii\,x^3$ addition to the harmonic potential. The corresponding Hamiltonian is
\PT symmetric, that is, symmetric under a simultaneous complex conjugation
and space reflection. Even though this is not quite obvious, such a Hamiltonian
has a real spectrum, as first conjectured by Bessis and Zinn-Justin (1992) and proven
in \cite{rDDB,rshin}. In this note, we show numerically that the ODM method,
suitably adapted to the problem, allows a precise determination of the ground
state energy. The convergence properties also give some information about the
analyticity  as a function of the strength of the $\ii\,x^3$ perturbation, in
particular excluding level crossing in  some region of the Riemann surface.

Another potential application of the method is the $\ii\phi^3$ quantum field
theory in two and three space dimensions, whose renormalization group functions
describe the Lee--Yang edge singularity of Ising-like statistical models
\cite{rLYFisher,rBBJ}, should longer perturbative series become available.

%
% The imaginary cubic potential
% 
\section{The imaginary cubic potential}
\label{ssLOBphiiv}

We here consider the spectrum of the simplest example of a \PT symmetric Hamiltonian,
\begin{equation}
H =- \frac12 \,\frac{\dd^2}{\dd x^2}+ \frac12 \, x^2+ 
\tfrac{\ii}{6} \, \sqrt{g} \; x^3\,,
\label{eODMHixiii}
\end{equation}
where $g$ is a real positive parameter. Indeed, the Hamiltonian  
is invariant under simultaneous complex conjugation and 
parity transformation $x \mapsto -x$. 
In this convention, the resonance energies of the cubic potential
have their branch cut along the negative real $g$ axis, where
the quantity $\ii \, \sqrt{g}$ becomes real and the particle
may tunnel through the ``cubic wall'' of the potential.
One then verifies that the perturbative expansion of the
energy eigenvalues $E_n$ for $g\to0$ contains only integer powers of $g$:
\begin{equation}
E_n(g)=\sum_{k=0}^\infty E_{n,k} g^k
\end{equation}
with real coefficients. Moreover, a steepest descent calculation of the path
 integral representation of the corresponding quantum partition function \cite{rLOBLip,rLOBgen,rZJLOreport} 
yields a large order behaviour of the form
\begin{equation}
\label{eLObehaviour}
E_{n,k} \mathop{\sim}_{k\to\infty} C \; (-1)^{k} \, k^b \,  A^{-k} \, k!\,,
\end{equation}
where $A=24/5$ is the instanton action. The constant $C$ and the half-integer
$b$ depend on $n$.  The series is divergent for all $g$ and when the expansion
parameter is not small, a summation of the perturbative expansion is
indispensable. Note that the sign oscillation with $k$ and some additional
considerations already suggest that the series is Borel summable (a property
proved in \cite{rCGM}) and thus that the spectrum, beyond perturbation theory,
is real.  This was initially the origin of the conjecture of Bessis and
Zinn-Justin, which was generalized to other \PT symmetric Hamiltonians
\cite{rBeBo,rBBrJ,rCMBender}.  More recently, Pad\'{e} summability was also
rigorously established \cite{rVGMMAM}. This proof confirms numerical
investigations based on the summation of the perturbative expansion by Pad\'{e}
approximants \cite{rBeWe}. Distributional Borel summability of the 
perturbation series to the complex resonance energies for negative coupling 
$g$ was proved in \cite{rCaliceti}.
 
In this note, we show by a combination of numerical and analytic arguments that
the ODM method is especially well adapted to summing the perturbative series,
converging for all values of the expansion parameter in the complex plane and,
in addition, in a region of the second sheet of the Riemann surface. It thus
also provides  additional information about the spectrum for $g$ complex on the
Riemann surface \cite{rHKWJ}.
 
However, before dealing with the quantum mechanics example, as a warm-up
exercise, we first consider the simple integral
\begin{equation}
Z(g)=\frac{1}{\sqrt{2\pi}}
\int_{-\infty}^{+\infty}
\dd x\,\exp\left(-\tfrac12 x^2-\tfrac{\ii}{6}\sqrt{g} x^3\right) , 
\end{equation}
whose expansion coefficients count the number of Feynman diagrams that appear
in the expansion of the partition function $\tr\ee^{-\beta H}$ corresponding to
the Hamiltonian \eqref{eODMHixiii}. 

%
% Figure 1
%
\begin{figure}[thb]
\includegraphics[width=1.0\linewidth]{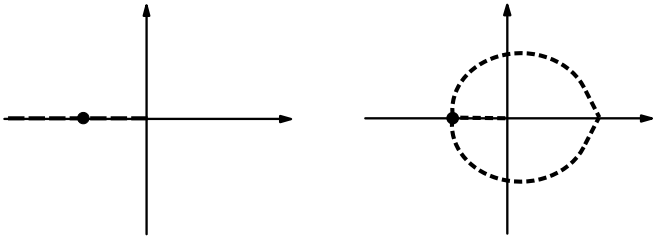}
\caption{\label{figodmi} Mapping: example of a function analytic in a cut-plane. 
The cut in the $g$-plane (l.h.s.) is mapped onto the contour 
$\bar\Gamma$ of the $\lambda$-plane (r.h.s.).}
\end{figure}
 
%
% Order-dependent mapping summation method
%
\section{Order-dependent mapping summation method} 
\label{ssODMdef}

%
% Method
%
\subsection{Method}

The ODM summation method \cite{rsezzin} is based on some {\it a priori} knowledge, 
or educated guess, of the analytic properties of the function that is expanded.
It applies both to convergent and divergent series, although it is mainly useful in
the latter case~\cite{rsezzin}.

Let $E(g)$ be an analytic function that has the Taylor series expansion 
\begin{equation}
E(g)=\sum_{\ell=0}^\infty E_\ell\, g^\ell \,,
\end{equation}
where the equality has to be understood in the sense of a
(possibly asymptotic) series expansion.
In what follows, we consider two functions analytic at least in a sector (as in
the example of Fig.~\ref{figodmi}) and only mappings $g\mapsto \lambda$ of the
form 
\begin{equation}
\label{eODMmapping} 
g = \rho \; \frac{\lambda}{(1-\lambda)^\alpha}\,, \qquad \alpha >1 \,,
\end{equation}
where  $\alpha$ has to be chosen in accordance with the analytic properties of
the function $E$ (a rational number in our examples) and $\rho$ is an
adjustable parameter. A general discussion of the method can be found in
Refs.~\cite{rsezzin,rZJODM}. An important property that singles out such a
mapping is the following: inverting the mapping, one finds that,
for $g \to \infty$, $\lambda$ approaches unity,
and $g^{-1/\alpha}$ has a regular expansion in powers of
$1 - \lambda$ (see the Appendix).

After the mapping \eqref{eODMmapping}, 
$E$ is given by a Taylor series in $\lambda$ of the form
\begin{equation}
\label{trunc}
E\bigl( g(\lambda) \bigr) = \sum_{\ell =0}^\infty P_\ell(\rho) \; \lambda^\ell,
\end{equation}
where the coefficients $P_\ell(\rho)$ are polynomials of degree $\ell$ in
$\rho$.  Since the result is formally independent of the parameter $\rho$, the
parameter can be chosen freely. At  $\rho$ fixed, the series in $\lambda$ is
still divergent, but it has been verified on a number of examples (all Borel
summable), and proved in certain cases \cite{rDuJo,rRGKKHS} that, by adjusting
$\rho$  order by order, one can devise a convergent algorithm.  The basic idea
is that  $\rho$ characterizes a ``local radius of convergence'' of the
divergent input series.  As the order of the mapping is increased, $\rho \to
0$. However, the mapping $g = g(\rho, \lambda)$ is constructed so that the
circle of convergence of order-dependent radius $\rho$ is always mapped onto a
finite domain of the $\lambda$ complex plane.  

At a risk of some oversimplification, we can describe the general paradigm of
the ODM summation as follows: One ``pretends'' that the divergent input series
is in fact a convergent series, but with a circle of convergence whose radius
gradually vanishes as the order of the perturbative expansion is increased.
Since both $\rho \to 0$ and also $|\lambda | < 1$, the direct expansion in
$\lambda$, with variable $\rho$, may then lead to a convergent output series
even for large $g$.

The $k$-th approximant  $E^{(k)}(g)$ is constructed in the following way: one
truncates the expansion~\eqref{trunc} at order $k$ and chooses $\rho = \rho_k$
as to cancel the last term, \textit{i.e.} $P_k(\rho_k) = 0$.  Because
$P_k(\rho)$ has $k$ roots (real or complex), in general one chooses for $\rho$
the largest possible root (in modulus) $\rho_k$  for which $P_k'(\rho)$ is
small (but see the more detailed discussion of Sec.~\ref{ssODMconvergence}). 
 
This leads to a sequence of approximants 
\begin{equation}
\label{ODMapprox}
E^{(k)}(g)=\sum_{\ell=0}^{k} P_\ell(\rho_k)\;\lambda^\ell(g,\rho_k) \quad 
\mbox{with}\quad P_k(\rho_k)=0\,,
\end{equation}
where $\lambda = \lambda(g,\rho)$ is obtained by inverting
the relation~\eqref{eODMmapping}.
In the case of convergent series, it is expected that $\rho_k$ has a
non-vanishing limit for $k\to\infty$.  By contrast, for divergent series it is
expected that $\rho_k$ goes to zero for large $k$ as

\begin{equation}
\rho_k= \calO\left( E_k^{-1/k}\right) \,.
\end{equation}
The intuitive idea here is that $\rho_k$ is proportional to a 
``local'' radius of convergence. 

The following remarks are in order.
(i) Alternatively, one can choose the largest roots $\rho_k$ of the polynomials
$P'_{k}(\rho )$  for which $P_k$ is small.  Other mixed criteria involving a
combination of $P_k$ and $P'_k$ can also be used. Indeed, the approximant is
not very sensitive to the precise value of $\rho_k$, within errors. Finally,
$P_{k+1}(\rho_k)\lambda^{k+1}$, as well as the values of $P_k(\rho)\lambda^k$ in the neighbourhood of $\rho_k$,  give  an order of magnitude of the error.
(ii) In the ODM method, the main task is to determine the sequence of $\rho_k$. 
Indeed, once the $\rho_k$ are known, for each value of $g$ the 
calculation reduces to inverting the mapping \eqref{eODMmapping} and simply 
summing the Taylor series in $\lambda$ to the relevant order.

%
% Optimal convergence analysis
%
\subsection{Optimal convergence analysis}
\label{ssODMconvergence}

We here give a heuristic analysis \cite{rsezzin} of the convergence of the ODM
method that shows how the convergence can be optimized. This will justify the
choice of the class of zeros of the polynomials $P_k$ or $P'_k$  and provide a
quantitative analysis of the corresponding convergence.

In this note, we consider only real  functions analytic in a 
cut plane with a cut along the real negative axis (see Fig.~\ref{figodmi}) 
and a Cauchy representation of the form  
\begin{equation}
E(g)= \frac{1}{\pi}\int^{0_-}_{-\infty}
\dd g' \frac{\Im E(g'+i0)}{g'-g}\,,
\end{equation}
up to possible subtractions to  ensure the convergence of the integral for $g\to-\infty$. For later purpose, we introduce the notation
\begin{equation}
\Delta(g)=\Im\, E(g+i0).
\end{equation}
Moreover, we assume that the function $\Delta$ asymptotically fulfils
\begin{equation}
\label{eDgasymptotic}
\Delta (g) \mathop{\propto}_{g\to 0_-} 
(-g)^{-b-1} \ee^{A/g}  \,,\qquad A>0\,.
\end{equation}
For the two examples discussed below, 
this exponential asymptotic form can be 
derived by steepest descent calculations \cite{rLOBLip,rLOBgen,rZJLOreport}.
The function $E(g)$ can be expanded in powers of $g$:
\begin{equation}
E(g)=\sum_{\ell = 0}^\infty E_\ell \, g^\ell \qquad \mbox{with} \qquad 
E_\ell = \frac{1}{\pi} \int_{-\infty}^{0_-} 
\frac{\dd g}{g^{\ell+1}} \; \Delta (g)\,.
\end{equation}
The assumption \eqref{eDgasymptotic} then leads to a 
large order behaviour of the form  
\begin{equation}
E_k\mathop{\propto}_{k\to\infty}
(-A)^{-k} \, \Gamma(k+b+1) \sim (-A)^{-k} \, k^{b}\, k!
\end{equation}
as assumed in Sec.~\ref{ssLOBphiiv}.
We now introduce the mapping \eqref{eODMmapping}, 
\begin{equation}
g=\rho \frac{\lambda}{(1-\lambda )^\alpha }\,.
\end{equation}
The cut on the real negative axis of the $g$-plane is then mapped onto the contour $\bar\Gamma$ (Fig.~\ref{figodmi}).

The Cauchy representation then can be written as
\begin{equation}\label{eCauchylambda}
E\bigl(g(\lambda )\bigr) = \frac{1}{2i\pi}\,\oint_{\bar\Gamma}\dd  \lambda' {E\bigl(g(\lambda')\bigr)\over\lambda'-\lambda}=
\frac{1}{\pi} \,
\oint_{\Gamma} \dd  \lambda'  
\frac{\Delta \bigl(g(\lambda ')\bigr)}{\lambda  - \lambda'} \,,
\end{equation}
where $\Gamma$ is the image of the upper-part of the cut on the real negative axis
under the mapping $\lambda = \lambda(g)$ at fixed $\rho$, 
with a segment of the real $\lambda$ negative axis, $1/(1-\alpha)<\lambda'<0$, 
and the complex contour $\Im(\bar\Gamma)>0$ ending at $\lambda=1$. 
The contours can be deformed if the function $E(g)$ has 
analyticity properties beyond the first Riemann sheet.

We expand, omitting the dependence on $\rho$ in $\lambda$,
\begin{equation}
\label{eEODMlambda}
E\bigl(g(\lambda )\bigr)=\sum_{\ell = 0}^\infty P_\ell(\rho)[\lambda (g)]^\ell 
\end{equation}
where the approximants are obtained by the truncated expansion
\begin{equation}
\label{eEODMlambdatrunc}
E\bigl(g(\lambda)\bigr) \approx\sum_{\ell = 0}^k P_\ell(\rho_k)[\lambda (g)]^\ell \,,
\end{equation}
and $\rho_k$ is chosen so that 
$P_k(\rho_k) \approx 0 $. Moreover, 
\begin{equation}
P_\ell(\rho)= \frac{1}{\pi}
\oint_{\Gamma} \frac{\dd\lambda}{\lambda^{\ell + 1}} \,
\Delta \bigl(g(\lambda  )\bigr)   .
\label{eODMPkgen} 
\end{equation}
For $k\to\infty$, the factor $\lambda^{-k-1}$ favours small values of 
$|\lambda|$, but for too small values of $\lambda$, the exponential decay of  
$\Delta\bigl(g(\lambda)\bigr)$ takes over. Thus, the 
remainder value of the polynomial $P_k(\rho_k)$ can be evaluated by the
steepest descent method. The ansatz  
\begin{equation}
\label{rhokasym}
\rho_k\sim R/k\,,\qquad R>0\,,  
\end{equation}
implies that the saddle-point values of $\lambda$ are independent of $k$
and $g(\lambda )\to0$. Thus,  $\Delta (g)$ can be replaced 
by its asymptotic form \eqref{eDgasymptotic} for $g\to0_-$, 
except for $g$ of order $1$ and thus $\lambda$ close to $1$. 
In what follows, we set  
\begin{equation}
\mu = \frac{R}{A}\,,
\end{equation}
since this is the only parameter 
(and it is independent of the normalization of $g$). 
The behaviour of $P_k(\rho_k)$,
with $\rho_k$ being given by Eq.~\eqref{rhokasym},
is then given by the leading saddle point contributions:
\begin{equation}
\label{eODMPkbehav}
P_{k}(\rho_k)= \calO(\ee^{k\sigma }) \quad \mbox{with} \quad
\sigma = \frac{1}{\lambda\mu }(1-\lambda )^\alpha - \ln \lambda \,.
\end{equation}
At a zero of $P_k(\rho_k)$, several leading contributions cancel but \textit{each contribution is expected to give an order of magnitude of the error}.\par
The saddle point equation is   
\begin{equation}
\label{eODMsaddle} 
\sigma'(\lambda)= -\frac{1}{\mu\lambda^2} 
\left[ (1-\lambda)^{\alpha-1} 
\bigl(1-(1-\alpha)\lambda\bigr) +\lambda\mu\right]=0\,.
\end{equation}
For $\mu$ large, the equation  has a unique  solution (for $\mu\to0$, $\lambda\sim-1/\mu$), which is real negative 
 and also exists for all values of $\mu$ with $ 1/(1-\alpha)<\lambda<0$.
Then we note that at this saddle point, as a function of $\mu$,
\begin{equation}
\frac{\dd \sigma}{\dd \mu} =
-\frac{(1-\lambda )^\alpha}{\lambda\mu^2} > 0\,,
\end{equation}
since the saddle point value of $\lambda$ is negative. 
This suggests decreasing $\mu$ as much as possible to improve the convergence,
and thus,  taking the zero $\rho_k$ with the smallest modulus. 
The exponential rate $\sigma$ corresponding to the saddle point vanishes for
\begin{equation}
\label{eODMsing} 
\frac{1}{\lambda\mu }(1-\lambda )^\alpha -\ln(-\lambda )=0\,,
\end{equation}
and this defines a special value $\mu_c$ of the parameter $\mu$. 
Some numerical values, obtained by combining
Eqs.~\eqref{eODMsaddle} and \eqref{eODMsing} are displayed in 
Table~\ref{tabODMsaddle}. 
For $\mu>\mu_c$, this contribution to $P_{k}(\rho_k)$ grows exponentially
while for $\mu<\mu_c$ it decreases exponentially.\par
%%%%%%%%%%%%%%%%% HERE

The further analysis then somewhat depends on whether $\alpha$ is smaller or
larger than $2$.  If $\alpha>2$, there exists a region on $\Gamma$ where $\sigma$ is positive and convergence is only possible if the initial integration contours $\Gamma$ or $\bar\Gamma$ can be deformed in what corresponds to the second sheet of the function $E(g)$. More generally, the contribution
of the contour from the neighbourhood of $\lambda=1$ plays an important role except, again, if it is possible to deform $\bar\Gamma$ in the first expression \eqref{eCauchylambda} in such a way that $\lambda=1$ is no longer on the contour but inside it. 

Two cases need to be distinguished:
 
(i)  One can decrease $\mu$ until other, complex, saddle points or maxima of
the modulus on the contour give contributions with the same modulus. Relevant
zeros of the polynomials $P_k$ and $P'_k$ then correspond to the  cancellation
between the different saddle point contributions. If then $\mu>\mu_c$, $\sigma$
is positive and the new approximants eventually also diverge and just yield a
better behaved asymptotic expansion. If $\mu<\mu_c$, 
$ \sigma$ is negative and the contributions to $P_k(\rho_k)$ decrease  exponentially with $k$
(this implies the possibility of contour deformation),
and, as we show later, the method converges for all values of the parameter
$g$. This also implies that the mapping~\eqref{eODMmapping} removes all singularities of
$E(g)$, obviously an exceptional situation but illustrated by the example of
Sec.~\ref{ssODMintxiii}.
 
(ii) Generically, the mapping  \eqref{eODMmapping}  does not cancel all
singularities of the function $E(g)$ and if it is possible to decrease $\mu$ up
to $\mu_c$, $\sigma$ vanishes and then other contributions of order at most
$\exp\left( C \, k^{1-1/\alpha}\right)$ (in particular coming from the
integration near $\lambda=1$) appear and the relevant zeros of $P_k$ and $P'_k$
correspond to the cancellation of these contributions. One can further decrease
$\mu$ but the convergence properties are unlikely to improve since contributions
of order $\exp\left( C \, k^{1-1/\alpha}\right)$ survive and can no longer be reduced.

Returning to the expansion \eqref{eEODMlambda}, at $g$ fixed, 
from the behaviour of $\rho_k$ one  infers
\begin{equation}
1-\lambda \sim \left(\frac{R}{kg}\right)^{1/\alpha} 
\end{equation}
and so
\begin{equation}\label{elambdafactor}
\lambda^k \sim \exp\left(-k^{1-1/\alpha} \; 
\left(\frac{R}{g}\right)^{1/\alpha}\right) \,.
\end{equation}
If the saddle point with $\lambda<0$ is a leading saddle point and
$\mu\ne\mu_c$ [case (i)], convergence is dominated by the behaviour of
$\ee^{k\sigma}$ and this irrespective of the value of $g$.

If the integral is dominated by contributions of order $1$ in the sense of case
(ii), $\mu$ must be chosen in the range $\mu<\mu_c$ to avoid exponential
divergence, but then to decrease the factor $\lambda^k$ one should choose $R$,
and thus $\mu$, as large as possible. This implies choosing $\mu=\mu_c$. 

In a generic situation, we then expect  the behaviour of the contributions to $P_k(\rho_k)$ to be dominated by a factor of the form
\begin{equation}
 \exp\left( C\,k^{1-1/\alpha}\right) \,,
\end{equation}
and the domain of convergence depends on the sign of the constant $C$.
For $C>0$, the domain of convergence is
\begin{equation}
|g| < R\, C^{-\alpha } \, [\cos({\rm arg}(g)/\alpha ) ]^\alpha.
\end{equation}
For $\alpha >2$, this domain extends beyond the first Riemann sheet and
requires analyticity of the function $E(g)$ in the corresponding domain.

For $C<0$, the domain of convergence is the union of the sector 
$|{\rm arg}(g)|<\pi\alpha /2$ and the domain 
\begin{equation}
|g| > R\,|C|^{-\alpha }[-\cos({\rm arg}(g)/\alpha ) ]^\alpha \,.
\end{equation}
Again for $\alpha >2$, this domain extends beyond the first Riemann sheet.

\begin{table*}
\caption{\label{tabODMsaddle} Values of $\mu_c$ and $\lambda_c$.}
\begin{tabular}{c@{\hspace*{0.2in}}@{\hspace*{0.2in}}c@{\hspace*{0.2in}}%
c@{\hspace*{0.2in}}c@{\hspace*{0.2in}}c@{\hspace*{0.2in}}c}
\hline
\hline
\rule[-2mm]{0mm}{6mm} $\alpha$ & 3/2 & 2 & 5/2 & 3 & 4 \\
\rule[-2mm]{0mm}{6mm}
$\mu_c$ & 4.031233504 & 4.466846120 & 4.895690188 & 5.3168634291 & 6.1359656420 \\
\rule[-2mm]{0mm}{6mm} $-\lambda_c$ & 
0.2429640300 & 0.2136524524 & 0.1896450439 & 0.1699396648 & 0.14003129119 \\
\hline
\hline
\end{tabular}
\end{table*}

%
% Example of the $\ii\,x^3$ integral
%
\section{Example of the $\bm{\ii\,x^3}$ integral}
\label{ssODMintxiii}

We now study, as an example, the integral
\begin{equation}
\label{eIntxcube}
Z(g)= \frac{1}{\sqrt{2\pi}}\,
\int_{-\infty}^{+\infty}\dd x\,
\exp\left( -\tfrac12 x^2- \tfrac{\ii}{6}\,\sqrt{g}\;x^3 \right)\,.
\end{equation}
The function has a divergent series expansion in powers of $g$,
\begin{equation}
Z(g)=1-\frac{5}{24}g+\frac{385}{1152}g^2+\calO(g^3) \,,
\end{equation}
and the coefficient $A $ in \eqref{eLObehaviour} is 
given by the non-trivial saddle point 
\begin{equation}
x+\tfrac{\ii}{2} \,\sqrt{g} \; x^2=0 \quad \Rightarrow \quad 
x=\frac{2\ii}{\sqrt{g}} \,,
\end{equation}
and so the saddle-point value of the integrand is
\begin{equation}
\ee^{A/g} \quad \mbox{with} \quad  A=\frac23\,.
\end{equation}
Moreover, the function $Z(g)$ has a convergent large $g$ expansion of the form
\begin{equation}\label{eZxiiiglarge}
Z(g)=g^{-1/6}\sum_{\ell=0}^\infty z_\ell \; g^{-\ell/3}\,,
\end{equation}
where the evaluation of the integral yields
\begin{equation}
\label{z0}
z_0=6^{1/3}\frac{\sqrt{2\pi}}{3\; \Gamma(\tfrac23)} =
1.1212331717419689582\ldots
\end{equation}
This suggests the mapping (see the Appendix) 
\begin{equation}
g=\rho \frac{\lambda}{(1-\lambda)^3} 
\end{equation}
and the introduction of the function
\begin{equation}
\phi(\lambda,\rho)=(1-\lambda)^{-1/2}Z\bigl(g(\lambda)\bigr).
\end{equation}
Setting in the integral $x=s\sqrt{1-\lambda }$, one obtains
%%%%%%%%%%%%%%%%%%%%%%% align???? array
\begin{align}
\phi(\lambda,\rho )=& \; \frac{1}{\sqrt{2\pi}}
\int_{-\infty}^\infty \dd s\,
\exp\left( -\tfrac12 s^2 + \tfrac12 \lambda s^2 -
\tfrac{\ii}{6}\,\sqrt{\lambda \rho} s^3 \right)
\nonumber\\[2ex]
=& \; \sum_{\ell=0}^\infty \phi_\ell \lambda^\ell
%\end{array}
\end{align}
where 
\begin{align}
\phi_k=& \; \frac{1}{\sqrt{2\pi}} \,
\frac{1}{2\ii\pi} \, \oint_{\bar\Gamma} \frac{\dd\lambda}{\lambda^{k+1}}
\nonumber\\[2ex]
& \; \times \int_{-\infty}^\infty \dd s\,
\exp\left( -\tfrac12 s^2 + \tfrac12 \lambda s^2 -
\tfrac{\ii}{6}\,\sqrt{\lambda \rho} s^3 \right) \,.
\end{align}
The first terms are
\begin{align}
\phi(\lambda,\rho )=& \; 1 + 
\left(\frac12-\frac{5}{24}\rho\right) \, \lambda 
\nonumber\\[2ex]
& \; + \left(\frac{3}{8} -\frac{35}{48}\rho +
\frac{385}{1152}\rho ^2\right)\lambda^2+\calO(\lambda^3)\,.
\end{align}
The asymptotic behaviour of $Z$  
for $g\to\infty$, 
\begin{equation}
Z(g)\sim z_0\, g^{-1/6}\, 
\end{equation}
is obtained for $\lambda\to1$,
\begin{equation}
Z(g)\sim g^{-1/6}\rho^{1/6}\phi(1,\rho).
\end{equation}
It follows
\begin{equation}
\phi(1,\rho)=  z_0\,\rho^{-1/6}.
\end{equation}
We use the following 
values of $\rho_k$ in order to calculate the 
first ODM approximants $\phi^{(k)}(1,\rho)$ 
to $\phi(1,\rho)$ according to Eq.~\eqref{ODMapprox},
but with the condition $P'_k(\rho_k) = 0$
(zeros of the derivative) except for $k=1$ 
where we choose the condition $P_1(\rho_1) = 0$. 
These read (for $k = 1,2,3,4$)
\begin{equation}
2.4\,, \quad 1.0909\,, \quad  0.7058\pm 0.1866\, \ii\, \quad 
0.5894\pm 0.2633\,\ii\,. 
\end{equation}
The approximants to $z_0$ defined
in Eq.~\eqref{z0} yield 
(we give only the real parts when $\rho_k$ is complex)
\begin{subequations}
\label{approxgood}
\begin{align}
\rho^{1/6}\phi^{(1)}(1,\rho)= & \; 1.15709373\,, \\
\rho^{1/6}\phi^{(2)}(1,\rho)= & \; 1.26825944\,, \\
\rho^{1/6}\phi^{(3)}(1,\rho)= & \; 1.18358984\,, \\  
\rho^{1/6}\phi^{(4)}(1,\rho)= & \; 1.08048484 \,.
\end{align}
\end{subequations}
Surprisingly, using only a minimum number of terms from the 
input series, a rather good approximation to the strong-coupling limit
is obtained.

\begin{table}
\caption{\label{tabODMv} Values of $\rho_k$ and errors in the ODM 
approximants in the limit $g\to\infty$. 
For the estimated convergence rate of the output series obtained 
by the ODM, $(\delta_k)^{1/k}$ should approach a constant as $k \to \infty$.}
\begin{tabular}{ccccccc}
\hline
\hline
\rule[-2mm]{0mm}{6mm}
$k$                & 5 & 10 & 15 & 20 & 25 & 30 \\
\hline
\rule[-3mm]{0mm}{9mm}
$\dfrac{k\,\Re(\rho_k)}{A}$   & 3.3612 & 4.3300 & 4.3450 & 4.4210 & 4.4954 & 4.5365 \\
\rule[-3mm]{0mm}{9mm}
$\dfrac{k\,\Im(\rho_k)}{A}$   & 0.7187 & 0      & 0      & 0.4544 & 0.2984 & 0.2260 \\
\rule[-2mm]{0mm}{6mm}
$(\delta_k)^{1/k}$ & 0.4910 & 0.6149 & 0.6480 & 0.6526 & 0.4972 & 0.6819 \\
\hline
\rule[-2mm]{0mm}{6mm}
$k$                & 35 & 40 & 45 & 50 & 55 & 60 \\
\hline
\rule[-3mm]{0mm}{9mm}
$\dfrac{k\,\Re(\rho_k)}{A}$   & 4.54721 & 4.5509 & 4.5580 & 4.5698 & 4.5840 & 4.5884 \\
\rule[-3mm]{0mm}{9mm}
$\dfrac{k\,\Im(\rho_k)}{A}$   & 0.1800  & 0.4522 & 0.3706 & 0.5774 & 0      & 0.4390 \\
\rule[-2mm]{0mm}{6mm}
$(\delta_k)^{1/k}$ & 0.7008  & 0.7064 & 0.7004 & 0.7019 & 0.7232 & 0.7323 \\
\hline
\hline
\end{tabular}
\end{table}

\begin{figure}[th!]
\includegraphics[width=0.5\linewidth]{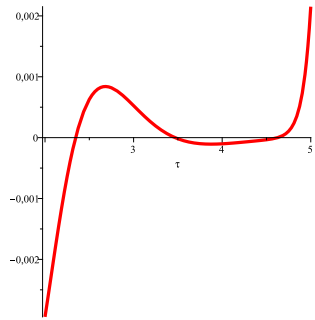}
\caption{\label{figodmii}
The ODM polynomial $P_k(\rho)$ as a function of 
the scaled variable $\tau=k\rho/A$ for $k=30$.
The model problem~\eqref{eIntxcube} is being studied.}
\end{figure}

\begin{figure}[th!]
\includegraphics[width=0.5\linewidth]{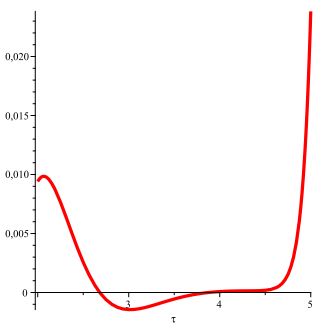}
\caption{\label{figodmiii}
The plot shows $AP'_k(\rho)/k$ as a function of $\tau=k\rho/A$ for $k=30$,
for the model problem~\eqref{eIntxcube}.}
\end{figure}

\begin{figure}[th!]
\includegraphics[width=0.5\linewidth]{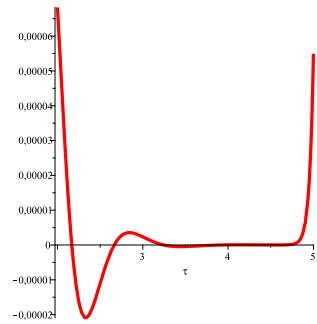}
\caption{\label{figodmiv}
The plot shows $P_k(\rho)$ as a function of $\tau=k\rho/A$ for $k=60$,
for the model problem~\eqref{eIntxcube}, demonstrating that a range 
where $P_k(\rho)$ is small persists in higher transformations orders.}
\end{figure}

\begin{figure}[th!]
\includegraphics[width=0.5\linewidth]{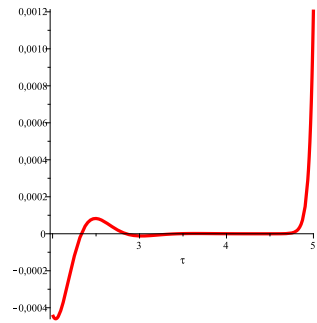}
\caption{\label{figodmv}
The plot shows $AP'_k(\rho)/k$ as a function of $\tau=k\rho/A$ for $k=60$,
again for the model problem~\eqref{eIntxcube}.
We demonstrate that the favourable region 
where both $P_k(\rho)$ and $P'_k(\rho)$ 
are small persists in higher transformations orders.}
\end{figure}

Figures \ref{figodmii},  \ref{figodmiii}, \ref{figodmiv} and
\ref{figodmv} show that for $P_k(\rho)$, in a rescaled variable 
$\tau = k \rho/A$, there  is an essentially
$k$-independent range where $P_k(\rho)$ and $P'_k(\rho)$ are small. This is
also the region in the neighbourhood of which the zeros of $P_k$ and $P'_k$ are
located. Quite generally, we choose
for $\rho_k$ the complex zeros of $P'_k(\rho)$ with the
largest real part in this range because they are easy to systematically identify.

The expected theoretical value, obtained by expressing that the real negative
and the two complex conjugate solutions of the saddle point equation
\eqref{eODMsaddle}, 
\begin{subequations}
\begin{align}
\lambda=& \; -0.1897168\ldots\,, \\
\lambda=& \; 0.8448584\ldots\pm \ii \, 1.386261\ldots \,,
\end{align}
\end{subequations}
yield contributions that have the same modulus (which allows for compensation)
is $\mu=R/A\approx 4.62987613\ldots$ and then the rate of convergence is
predicted to be of the form $0.775^k$  from the corresponding estimate
\eqref{eODMPkbehav}.

\begin{figure}[th!]
\includegraphics[width=0.5\linewidth]{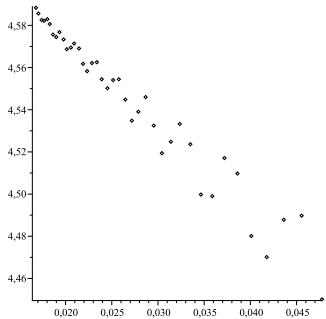}
\caption{\label{figodmvi}
$k\,\Re(\rho_k)/A$, odd-even averaged, as a function of $1/k$,
for the model problem~\eqref{eIntxcube}.
The plot illustrates how the asymptotic limit given 
in Eq.~\eqref{lim465} is approached for $k \to \infty$.}
\end{figure}

The numerical investigation for
$g\to\infty $ extrapolated to $k=\infty$ leads to (see Table~\ref{tabODMv}
and Fig.~\ref{figodmvi}) 
\begin{equation}
\label{lim465}
\lim_{k\to\infty} k \, \Re(\rho_k)/A = 4.65\pm0.03\,, 
\end{equation}
yielding a value of $\mu$ significantly smaller than the
value $5.31\ldots$ of Table~\ref{tabODMsaddle} in section
\ref{ssODMconvergence} and consistent with the theoretical expectation.

To calculate the ODM approximants, we fit $\rho_k$ combining the theoretical asymptotic form and the calculated values of $\rho_k$ in the range $k\le60$. This is achieved by adding a slightly \textit{ad hoc}\/ small correction term to the asymptotic formula.

\begin{figure}[th!]
\includegraphics[width=0.5\linewidth]{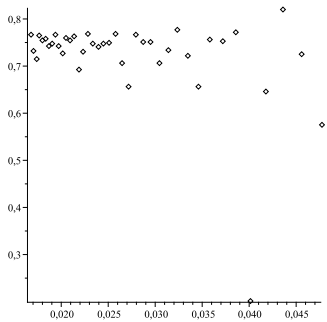}
\caption{\label{figodmvii}
The plot shows $(\delta_k/\delta_{19})^{1/(k-19)}$ as a function of $1/k$.
We illustrate the asymptotic rate of convergence 
of the ODM approximants to the strong-coupling limit
for the model problem~\eqref{eIntxcube}, in higher transformation
orders beyond $k = 19$.}
\end{figure}

For $g\to\infty$, we define the following quantity which estimates 
how well the asymptotic strong-coupling limit is approximated by the 
ODM approximants of order $k$,
\begin{equation}
\delta_k = 
\left| \frac{\left.\Re(Z(g))\,g^{1/6}\right|_{\rm ODM,k}}%
{\left. Z(g) \, g^{1/6} \right|_{\rm exact}} - 1
\right|\,.
\end{equation}
Some data are then displayed in Table~\ref{tabODMv} and 
Fig.~\ref{figodmvii}. From a fit to the numerical data
in Fig.~\ref{figodmvii}, One infers a geometric convergence of the form
$(0.78\pm0.02)^k$ quite consistent with the theoretical prediction.  When $g$
is finite and for $k\to\infty$, the additional factor coming from $\lambda^k$
has the form (see Sec.~\ref{ssODMconvergence})
\begin{equation}
\lambda^k  \sim \exp\left(-\frac{k^{3/5}R^{2/5}}{g^{2/5}} \right).
\end{equation}
This factor can never cancel the geometric convergence factor and, thus, the ODM
approximants converge on the whole Riemann surface. 

%
% Quantum mechanics
%
\section{Quantum mechanics}
\label{cubic}

%
% Strong coupling limit of the cubic potential
%
\subsection{Strong coupling limit of the cubic potential}

We now consider the spectrum of the Hamiltonian \eqref{eODMHixiii}, 
which can be obtained by solving the time-independent Schr\"{o}dinger equation
\begin{equation}
\label{eschrodingerxiii}
-\tfrac12 \psi''(x)+
\left(\tfrac12 x^2 +
\tfrac{\ii}{6} \, \sqrt{g}\; x^3
\right)\psi(x)=E \, \psi(x)
\end{equation}
with appropriate boundary conditions. The path integral representation of the
corresponding partition function has a structure that bears some analogy with
the integral \eqref{eIntxcube} discussed in Sec.~\ref{ssODMintxiii}.

The spectrum
has an expansion in integer powers of $g$ with real coefficients. For example,
for the ground state
\begin{equation}
E_0(g)=\frac12 + \frac{11}{216} \, g - \frac{155}{384} \,g^2 + \calO(g^3).
\end{equation}
Instanton calculus, based on a steepest descent evaluation of the corresponding
path integral, allows us to calculate the classical action relevant for the 
large order behaviour~\eqref{eLObehaviour}. One finds
\begin{equation}
A=\frac{24}{5}\,.
\end{equation}
For $g$ large, from the scaling properties of Eq.~\eqref{eschrodingerxiii}, one
infers that the ground state energy $E_0 = E_0(g)$,
as a function of $g$, for large $g$,  has an expansion of the form
\begin{equation}\label{eExiiilargeg}
E_0=g^{1/5}\sum_{\ell=0}^\infty \epsilon_\ell \, g^{-2\ell/5}.
\end{equation}
This suggests the mapping (see the Appendix)
\begin{equation}
g = \rho \frac{\lambda}{(1-\lambda)^{5/2}}\,,\quad 
E_0 = (1-\lambda)^{-1/2}\phi(\lambda,\rho) \,.
\end{equation}
In what follows, we again first concentrate on the behaviour of $E_0(g)$ for
$g\to\infty$, which is related to $\phi(1,\rho)$. More precisely, taking the
strong coupling limit $\lambda \to 1$, one finds
\begin{equation}
\phi(1,\rho)=\rho^{1/5}\epsilon_0\,.
\end{equation}

\begin{figure}[th!]
\includegraphics[width=0.5\linewidth]{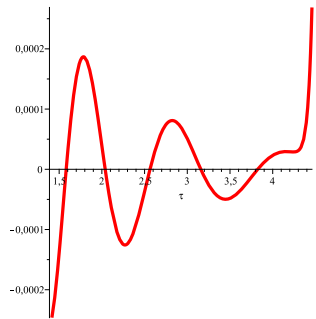}
\caption{\label{figodmviii} We show 
$P_k(\rho)$ as a function of $\tau=k\rho/A$ for 
the ODM approximant of order $k=55$ to the ground state energy 
of the cubic Hamiltonian~\eqref{eschrodingerxiii}.}
\end{figure}

\begin{figure}[th!]
\includegraphics[width=0.5\linewidth]{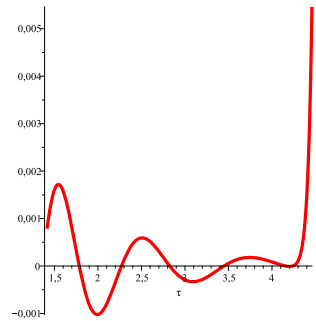}
\caption{\label{figodmix} 
The plot displays $AP'_k(\rho)/k$ as a function of 
the scaled variable $\tau=k\rho/A$ for 
the ODM approximant of transformation order $k=55$
to the ground state energy
of the cubic Hamiltonian~\eqref{eschrodingerxiii}.}
\end{figure}

We now proceed as for the model problem discussed in Sec.~\ref{ssODMintxiii}.
Figures~\ref{figodmviii} and \ref{figodmix}  show in a rescaled
variable, that there is a range, which is $k$-independent up to corrections
of order $k^{-2/5}$, where $P_k(\rho)$ and its derivative are small. This is
also the region in the neighbourhood of which the zeros of $P_k$ and $P'_k$ are
located. We have first determined the complex zeros $\rho_k$ of $P'_k(\rho)$
with the largest real part in this range. Again, these zeros have been chosen because it is easy to identify them systematically.  

A fit of the numerical data for orders $k\le 55$  and an extrapolation to $k\to\infty$ then yields
\begin{equation}
\label{eODMDimuxiii}
\frac{k\Re(\rho_k)}{A} =
5.5\pm0.2 - \frac{6.0\pm0.5}{k^{2/5}}\,,
\end{equation}
and thus
\begin{equation}
\label{mu545}
\mu=\frac{R}{A} =5.5\pm0.2\,, 
\end{equation}
to be compared with the expected maximum value $\mu=4.895\ldots $ of Table
\ref{tabODMsaddle}.

\begin{figure}[th!]
\includegraphics[width=0.5\linewidth]{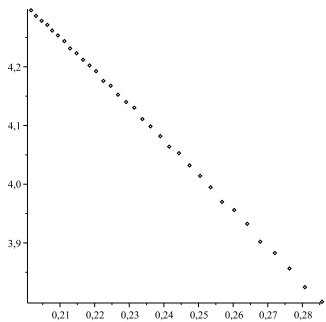}
\caption{\label{figodmx} $k\Re(\rho_k)/A$ as a function of $1/k^{2/5}$. 
The range for the abscissa is the interval $(0.20, 0.29)$,
and an extrapolation of the data to $k \to \infty$ is compatible
with Eq.~\eqref{eODMDimuxiii}.
%{\bf However, there seems to be a problem with the 
%numerical value of the constant coefficient ($4.3$ against $4.8$).}
}
\end{figure}

\begin{figure}[th!]
\includegraphics[width=0.5\linewidth]{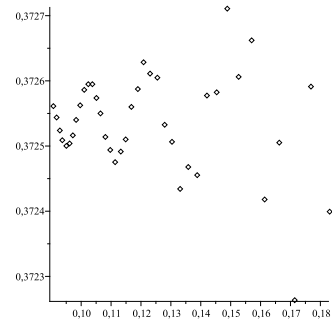}
\caption{\label{figodmxi} 
$E_{0,k}$, even-odd averaged, as a function of $1/k^{2/5}$.
As $k \to \infty$, the energy converges---under oscillations---to 
the point indicated as $1/k^{2/5} \to 0$.
The plot illustrates the findings of Eqs.~\eqref{trueEN} and~\eqref{trueER}.} 
\end{figure}

The numerical relative rate of convergence is compatible with an exponential of the form
$0.05 \times 0.94^k $ and yields the estimate 
\begin{equation}
\frac{E_0}{g^{1/5}}
\mathop{\rightarrow}_{g\to\infty}
0.3727\pm0.0003\,.
\end{equation}
However, the
value of $\mu$  would lead eventually to a divergence of order $1.2^k$. This
indicates that the form \eqref{eODMDimuxiii} cannot be used asymptotically. If we
take into account the expected value $\mu=4.895\ldots $ and again fit
$\Re(\rho_k)$, we find 
\begin{equation}
\label{trueK}
\frac{k \rho_k}{A} = 4.895690188- \frac{3.02}{k^{2/5}}\,,
\end{equation}
a form we use in the remaining calculations. With this form, we indeed find a
smoother convergence with
\begin{equation}
\label{trueEN}
\frac{E_0}{g^{1/5}} \mathop{\rightarrow}_{g\to\infty}0.37255\pm0.00004\,,
\end{equation}
at order $55$ to be
compared with the value obtained by solving the Schr\"{o}dinger equation
$E_0=0.372545791$ \cite{rUJJZJxiii}.

The errors can be fitted by the form
\begin{align}
\label{trueER}
& \tfrac12 (E_{0,k}+E_{0,k-1}) = E_{0\,{\rm as.}} \\
& - 0.00092\, \exp[-0.292\,k^{3/5}] \cos(4.02\,(k^{3/5}-1.20)),
\nonumber
\end{align}
where the odd-even averaging improves the fit of the error
(and slightly the convergence) by cancelling some subleading oscillatory correction. The form of the oscillatory factor in \eqref{trueER} is just a simple guess
(see Fig.~\ref{figodmxi}).

%
% Finite value of the coupling parameter
%
\subsection{Finite value of the coupling parameter}
 
For $g$ finite, the additional  factor \eqref{elambdafactor}  governing the 
convergence of the ODM approximants in transformation
order $k$  is  
\begin{equation}\label{eODMQMlambdak}
\lambda^k\sim \exp\left(-\frac{R^{2/5} \, k^{3/5}}{g^{2/5}} \right)
\end{equation}
with 
\begin{equation}
 R^{2/5}\approx 3.535\ldots\,.
\end{equation}
As a check, we have
summed the series for $g=5$. The result is 
\begin{equation}
E_0(5)=0.60168393320519(6), 
\end{equation}
and the convergence rate of the order of $\ee^{-2.10\,k^{3/5}}$. Comparing with
the convergence for $g\to\infty$, one finds a variation of the coefficient of
$k^{3/5}$ of $1.81$ to be compared with $3.5/g^{2/5}=1.86$.

For $g=1$, the
result obtained from a numerical solution of the Schr\"{o}dinger equation yields
\begin{equation}
E_0(1)=0.53078175930417667113556\,,
\end{equation}
with all digits being significant.
The ODM at order 55 yields (we give its explicit numerical
value for illustration)
\begin{equation}
E_0(1)=0.53078175930417667113565\,.
\end{equation}
The analysis of the convergence is
consistent with the theoretical form with a coefficient of $k^{3/5}$ equal to
$3.85$ compared to the theoretical expectation of $3.82$.

Finally, to
compare with the Pad\'{e} 
summation of Ref.~\cite{rBeWe}, we give  results for $g=288/49$,
which corresponds to $\lambda^2=1/7$ in Ref.~\cite{rBeWe}, %% The authors
%wanted to prove the Stieltljes property, this works only for the function
\begin{align}
k= 54\,,\quad E_0(\tfrac{288}{49})=& \; 0.6127381063888\,,\nonumber\\
\Delta E_0(\tfrac{288}{49})=& \; 5.524167213054\,,\\
k= 55\,,\quad E_0(\tfrac{288}{49})=& \; 0.6127381063891\,\nonumber\\
\Delta E_0(\tfrac{288}{49})=& \; 5.524167213068\,.
\end{align}
where $\Delta E_0= 49 \, (E_0-\tfrac12)$. At order 54, in Ref.~\cite{rBeWe} one finds 
$\Delta E_0 = 5.5241683$ and, at order 192, 
$\Delta E_0 = 5.52416721306031$.  The theoretical formula for
the ODM methods predicts an error for $\Delta E$ of the order of a few times
$10^{-22}$ for $k=192$, confirming the improved convergence achievable by use
of the ODM. For completeness, we have verified that 
$\Delta E_0 = 5.5241672130602221133\dots$ by numerical solution of the 
Schr\"{o}dinger equation.

%
% The imaginary part on the real negative axis
%
\subsection{The imaginary part on the real negative axis}
\label{complex}

\begin{figure}[th!]
\includegraphics[width=0.8\linewidth]{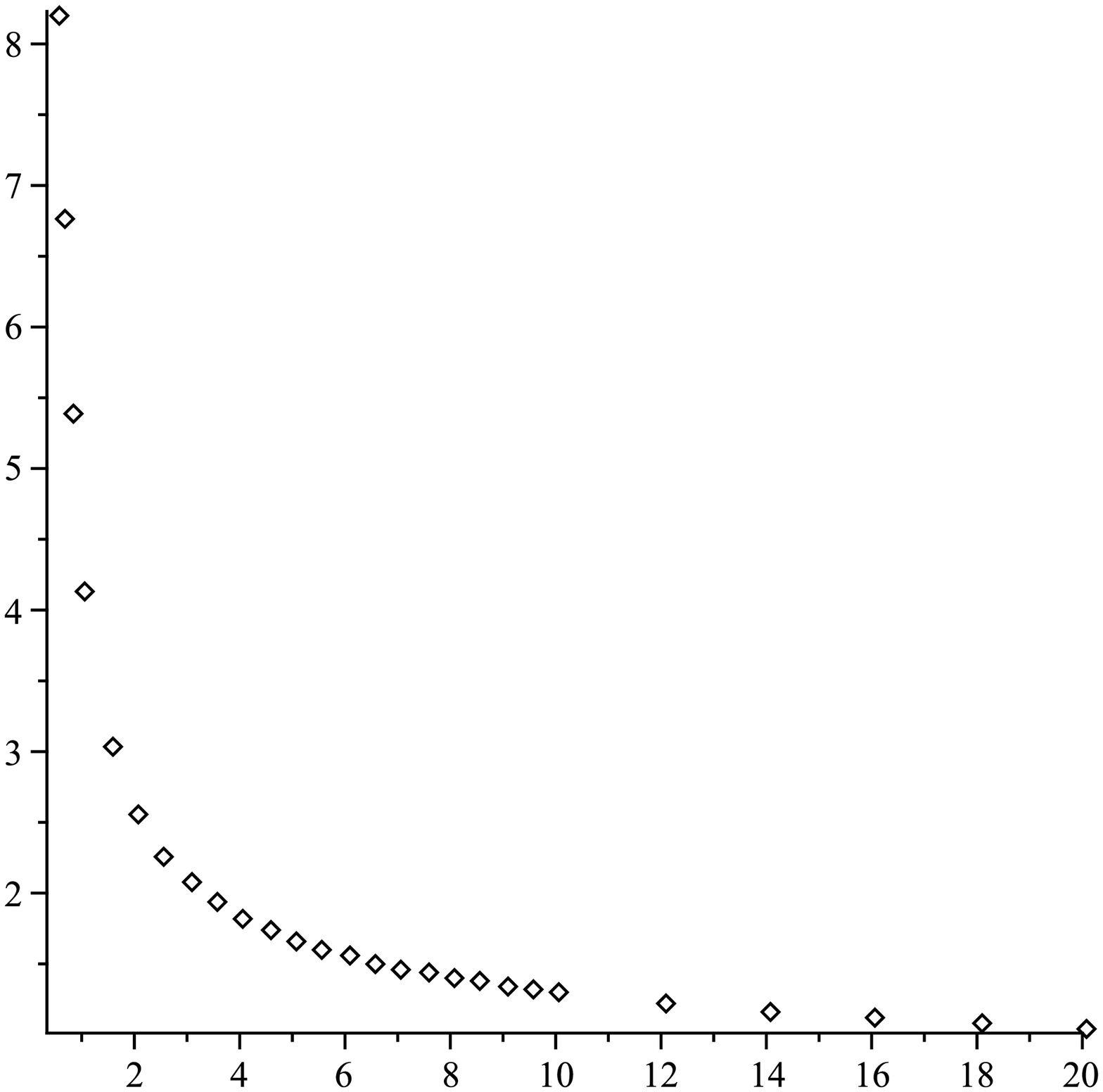}
\caption{\label{figodmxii} 
The function $-\ln\bigl(\Im E_0(g)\bigr)$ as a function of $-g$.} 
\end{figure}

The ODM method converges on the real negative axis and thus the Stieltjes
property~\cite{rVGMMAM}, that is, the positivity of the discontinuity,  can be
verified. For $g\to 0_-$ with $\Im(g) >0$, the imaginary part is known
analytically from semi-classical considerations (instantons) \cite{rUJJZJxiii}:
\begin{equation}
\Im E_0(g) \mathop{=}_{g\to 0_-} 
\frac{6}{\sqrt{\pi}} 
\frac{\ee^{24/5g}}{\sqrt{-g}} \, 
\left(1+\frac{169}{576} \,g + \calO(g^2)\right) \,,
\end{equation}
where we recall that we are using the cubic Hamiltonian 
here in the convention~\eqref{eODMHixiii}.
For $|g|\to\infty$ and $\Im(g) =0_+$, one finds
\begin{equation}
E_0(g \to -\infty + \ii \,0)\sim (0.301396+\ii \, 0.218977)\,|g|^{1/5} \,. 
\end{equation}
Above the negative real axis, the 
antiresonance energy has a positive imaginary part.
For $g=-21.6$, at order 55 we find $E_0=0.5540544 +0.351399\ii$ to be compared with the value
$E_0=0.554 053 519 + 0.351 401 778 \ii$ given in \cite{rUJJZJxiii}.
Similarly, for $g=-5+\ii 0$, $E_0=0.4338906+ 0.1838582\ii$, 
for $g=-1+\ii 0$, $E_0=0.4425200442+0.015517927\ii$. 
In fact, the imaginary part is a simple positive decreasing function providing a smooth interpolation between the two asymptotic forms (see Fig.~\ref{figodmxii}).
% Convergence and singularities
%
\subsection{Convergence and singularities}

Combining Eqs.~\eqref{trueER} and \eqref{eODMQMlambdak}, one finds that the convergence domain is  given by
\begin{equation}
\Re(g^{-2/5})>-0.0826\,.
\end{equation}
This contains the sector $|{\rm arg}\,g|<5\pi/4$, that is, the whole Riemann
sheet and a sector of the second sheet. Moreover, in the neighbourhood of
$g=\infty$, one finds the additional domain
%{\bf does the prefactor have an analytic representation?} Perhaps but it %has yet to be found
%
\begin{equation}
|g| > 510.0\; [-\cos(2{\rm arg}\,g/5)]^{5/2}.
\end{equation}
In the union of these two domains, the function $E_0(g)$ is analytic and thus
free of level-crossing singularities. In particular, the expansion
\eqref{eExiiilargeg} has about $0.0826$ as a radius of convergence.

%
% Conclusions
%
\section{Conclusions}
\label{conclu}

In this article, we have explored the properties of the
order-dependent mapping (ODM) for the summation of the divergent
series originating from non-Hermitian Hamiltonians.
In these cases, the resulting resonance energies
may be complex, and their behaviour may be studied 
as a function of the complex coupling parameter $g$.
In Sec.~\ref{ssLOBphiiv}, we discuss the motivation for the
current study and the two physical problems treated here, 
which are the cubic anharmonic oscillator in quantum mechanics,
and a model problem
that counts the number of Feynman diagrams that occur in the 
perturbative expansion of the partition function of the 
cubic anharmonic oscillator.

In Sec.~\ref{ssODMdef}, we recall some important properties of 
the ODM summation method, and in particular, we describe
how the saddle-point approximation helps in determining
asymptotic estimates for the parameter $\rho_k$ that are needed
in order to construct the ODM in higher orders, for the 
convergence rate of the resulting approximants in the strong
coupling limit, and for the analyticity domains of the 
approximants as the complex phase of the coupling parameter is varied.

In Sec.~\ref{ssODMintxiii}, we discuss the model problem
defined in Eq.~\eqref{ssODMintxiii} which is a simple
integral that counts Feynman diagrams in a cubic theory
and gives rise to a divergent series.
We demonstrate that good approximations to the 
strong coupling limit can be obtained on the basis of 
few perturbative terms [see Eq.~\eqref{approxgood}].
In Sec.~\ref{cubic}, the analysis is first carried over
to the \PT symmetric cubic potential, with similarly encouraging
numerical results, before demonstrating the applicability 
of the method to manifestly complex resonance energies
(see Sec.~\ref{complex}).

The basic idea of the ODM is that 
one  pretends  that the divergent input series,
which is to be summed,   is in
fact a convergent series, but with a circle of convergence
whose radius gradually
vanishes as the order of the perturbative expansion is increased. 
This algorithm, together with the representation of 
the coupling parameter $g$ given in Eq.~\eqref{eODMmapping},
leads to a very rapid convergence rate of the ODM transforms,
in part because the double expansion in $\rho$ and 
$\lambda$ implied by Eq.~\eqref{eODMmapping} represents
$g$ as a function of two variables $\rho$ and $\lambda$ 
whose modulus is bounded
by unity for approximants constructed in the strong coupling limit.
If one additionally makes a suitable change of variable (and function),
in order to incorporate the information about the strong coupling asymptotic expansion  (see the Appendix), then one may approximate the 
strong coupling behaviour using only very few perturbative input data.

%
% Acknowledgments
%
\section*{Acknowledgments}

U.D.J.~acknowledges support by a Grant from the
Missouri Research Board and
by the National Science Foundation (Grant PHY--8555454). J.~Z.-J.~also gratefully acknowledges CERN's hospitality, where this work was completed.

\appendix

%
% Parameters in the ODM
%
\section{Parameters in the ODM}

\label{paramodm}

Let the strong-coupling asymptotic expansion 
for the quantity $E(g)$ be known,
\begin{equation}\label{eElargegexpansion}
E(g) = g^\beta \sum_{n=0}\epsilon_n g^{-n/\alpha} \,, \qquad g \to \infty\,,
\end{equation}
a property shared by the two examples we have discussed here (Eqs.~\eqref{eZxiiiglarge},\eqref{eExiiilargeg}) but also by the quartic anharmonic oscillator \cite{rsezzin} and all $x^N$ perturbations to the quantum harmonic oscillator. 
We consider the conformal mapping $g = \rho \lambda \, (1 - \lambda)^{-\alpha}$. This transformation maps the real positive $g$-axis onto the finite $\lambda$ interval $[0,1]$.\par
For  $g\to\infty$, $\lambda\to1$ and $g$ has an expansion at $\lambda=1$ of the form
\begin{equation}
g^{-1/\alpha}= \sum_{n=0 } \Lambda_n(1-\lambda)^{n+1}
\end{equation}
with $\Lambda_0=\rho^{-1/\alpha}$. 
%
%\begin{equation}
%g \sim \rho \, \frac{1}{(1 - \lambda)^\alpha} \,, \qquad g \to \infty\,.
%\end{equation}
%
The function 
\begin{equation}
\phi(\lambda) = (1-\lambda)^{\alpha\beta}   E\bigl(g(\lambda)\bigr)  
\end{equation}
has then a  Taylor series expansion  at $\lambda=0$,
\begin{equation}
\phi(\lambda)=\sum_{n=0}\phi_n\lambda^n, 
\end{equation}
 as well as at $\lambda=1$,
\begin{equation}
\phi(\lambda)=\sum_{n=0}\varphi_n(1-\lambda)^n 
\end{equation}
with $\varphi_0=\epsilon_0\rho^\beta $, where $\epsilon_0$ is the coefficient defined in \eqref{eElargegexpansion}. This last property explains, to a large extent, the good convergence of the method even for $g\to\infty$.


\begin{thebibliography}{10}
\expandafter\ifx\csname url\endcsname\relax
  \def\url#1{\texttt{#1}}\fi
\expandafter\ifx\csname urlprefix\endcsname\relax\def\urlprefix{URL }\fi
\expandafter\ifx\csname href\endcsname\relax
  \def\href#1#2{#2} \def\path#1{#1}\fi

\bibitem{rsezzin} R. Seznec, J. Zinn-Justin, \textit{Summation of divergent
series by order dependent mappings: Application to the anharmonic oscillator
and critical exponents in field theory }, J. Math. Phys. 20 (1979)
1398--1408.

\bibitem{rZJODM} J.~Zinn-Justin, \textit{Summation of divergent series:
Order-depen\-dent mapping}, arXiv:1001.0675 [math-ph].

\bibitem{rLGZJ} J. C. Le Guillou and J. Zinn-Justin, \textit{The hydrogen atom in
strong magnetic fields: summation of the weak field series expansion}, {\it
Ann. Phys. (N.Y.)} 147 (1983), 57-84.

\bibitem{rRGZJ} R. Guida and J. Zinn-Justin, \textit{3D Ising model: the scaling
equation of state}, {\it Nucl. Phys.} B489 (1997) 626-652.

\bibitem{rJLKMPRR} J. L. Kneur, M.B. Pinto, R.O. Ramos,  
\textit{Asymptotically
improved convergence of optimized perturbation theory in the Bose-Einstein
condensation problem}, Phys. Rev. A 68  (2003) 043615.

\bibitem{rKPR} J. L. Kneur, M. B. Pinto, R. O. Ramos,
\textit{Critical and tricritical
points for the massless 2D Gross--Neveu model beyond large $N$}, Phys. Rev. D 74
(2006) 125020.

\bibitem{rKPRS} J. L. Kneur, M. B. Pinto, R. O. Ramos, E. Staudt, 
\textit{Emergence of
tricritical point and liquid-gas phase in the massless 2+1 dimensional
Gross--Neveu model}, Phys. Rev. D 76  (2007)  045020.

\bibitem{rDuJo} A. Duncan, H. F. Jones, 
\textit{Convergence proof for optimized $\delta
$ expansion: Anharmonic oscillator}, Phys. Rev. D 47 (1993) 2560-2572.

\bibitem{rRGKKHS} R. Guida, K. Konishi, H. Suzuki, 
\textit{Improved Convergence Proof of
the Delta Expansion and Order Dependent Mappings}, Ann. Phys.  249 (1996)
109-145.

\bibitem{rWECas} W. E. Caswell, \textit{Accurate energy levels for the anharmonic
oscillator and a summable series for the double-well potential in perturbation
theory}, Ann. Phys.  123 (1979) 153-184.

\bibitem{rDDB} P. Dorey, C. Dunning and R. Tateo, \textit{Spectral
equivalences, Bethe ansatz equations, and reality properties in PT-symmetric
quantum mechanics,} J. Phys. A 34 (2001) L391; ibid. 34 (2001) 5679-5704.

\bibitem{rshin} K. C. Shin, \textit{On the reality of eigenvalues for a class of
PT-symmetric oscillators,} Commun. Math.  Phys. 229 (2002) 543-564.

\bibitem{rLYFisher} M. E. Fisher, \textit{Yang--Lee Edge Singularity and
$\phi^3$ Field Theory}, Phys. Rev. Lett. 40 (1978) 1610-1613.

\bibitem{rBBJ} C. M. Bender, D. C. Brody and H. F. Jones, \textit{Scalar Quantum
Field Theory with Cubic Interaction}, Phys. Rev. Lett. 93 (2004) 251601.

\bibitem{rLOBLip} L. N. Lipatov, \textit{Divergence of the perturbation-theory
series and pseudoparticles},  JETP Lett. 25 (1977) 104-107;
\textit{Divergence of the perturbation-theory series and the quasi-classical
theory},  Sov. Phys. JETP  45 (1977) 216-223.

\bibitem{rLOBgen} E. Br\'{e}zin, J. C. Le Guillou, J. Zinn-Justin,
\textit{Perturbation theory at large order. I. The $\varphi^N$ interaction},
Phys. Rev. D 15 (1977) 1544-1557.

\bibitem{rZJLOreport} J.~Zinn-Justin, \textit{Perturbation series at large
orders in quantum mechanics and field theories: application to the problem of
resummation}, Phys.~Rep. 70 (1981) 109--167.

\bibitem{rCGM} E. Caliceti, S. Graffi, and M. Maioli, \textit{Perturbation
theory of odd anharmonic oscillators,} Comm. Math.  Phys. 75 (1980)  51-66.

\bibitem{rBeBo} C. M. Bender and S. Boettcher, \textit{Real spectra in
non-hermitian Hamiltonian having PT symmetry}, Phys. Rev. Lett.  80 (1998)
5243.

\bibitem{rBBrJ} C. M. Bender, D. C. Brody, and H. F. Jones, \textit{Complex
Extension of Quantum Mechanics}, Phys.  Rev. Lett. 89, 270402 (2002) and Am. J.
Phys. 71, 1095 (2003).

\bibitem{rCMBender} C. M. Bender, \textit{Making sense of non-Hermitian
Hamiltonians}, Rep. Prog. Phys. 70 (2007) 947-1018.

\bibitem{rVGMMAM} V. Grecchi, M. Maioli and A. Martinez, \textit{Padé
summability of the cubic oscillator}, J. Phys. A: Math. Theor. 42 (2009) 425208.

\bibitem{rBeWe} C. M. Bender, E. J. Weniger, \textit{Numerical evidence that the
perturbation expansion for a non-Hermitian PT-symmetric Hamiltonian is
Stieltjes}, J. Math. Phys. 42 (2001) 2167-2183.

\bibitem{rCaliceti} E. Caliceti, \textit{Distributional \relax{B}orel 
summability of odd anharmonic oscillators},
J. Phys. A 33 (2000) 3753-3770.

\bibitem{rHKWJ} H. Kleinert, W. Janke, \textit{Convergence behavior of variational
perturbation expansion --- A method for locating Bender-Wu singularities}, Phys.
Lett. A 206 (1995) 283-289.

\bibitem{rUJASJZJ} U. D. Jentschura, A. Surzhykov, J. Zinn-Justin,
\textit{Generalized nonanalytic expansions, PT-symmetry and large order
formulas for odd anharmonic oscillators,} SIGMA  5 (2009) 005.

\bibitem{rUJJZJxiii} U. D. Jentschura, J. Zinn-Justin, \textit{ Calculation of
the Characteristic Functions of Anharmonic Oscillators}, ArXiv:1001.4313
[math-ph].

\end{thebibliography}
\end{document}